\documentstyle[aps,pre]{revtex}

   \font\tenmsb=msbm10 scaled\magstep 1
   \font\sevenmsb=msbm7 scaled \magstep 1
   \font\faivemsb=msbm5 scaled \magstep 1
\newfam\msbfam
      \textfont\msbfam=\tenmsb
      \scriptfont\msbfam=\sevenmsb
      \scriptscriptfont\msbfam=\faivemsb
\def\Bbb#1{{\fam\msbfam #1}}
\font\tengothic=eufm10 scaled\magstep 1
\font\sevengothic=eufm7 scaled\magstep 1
\newfam\gothicfam
      \textfont\gothicfam=\tengothic
      \scriptfont\gothicfam=\sevengothic

\newcommand{\be}{\begin{equation}}
\newcommand{\ee}{\end{equation}}
\newcommand{\dlt}{\delta}
\newcommand{\sgm}{\sigma}
\newcommand{\Gm}{\Gamma}
\newcommand{\gm}{\gamma}
\newcommand{\om}{\omega}
\newcommand{\br}{{\bf r}}
\newcommand{\al}{\alpha}
\newcommand{\ra}{\rightarrow}

\begin{document}

\draft

\title{Irreversibility of time for quasi-isolated systems}
\author{V.I. Yukalov}

\address{Bogolubov Laboratory of Theoretical Physics \\
Joint Institute for Nuclear Research, 141980 Dubna, Russia \\
and \\
Research Center for Optics and Photonics \\
Universidade de S\~ao Paulo, S\~ao Carlos, S\~ao Paulo 13560-970, Brazil}

\maketitle

\vskip 1cm

\begin{abstract}

A physical system is called quasi-isolated if it subject to small random
uncontrollable perturbations. Such a system is, in general, stochastically
unstable. Moreover, its phase-space volume at asymptotically large time
expands. This can be described by considering the local expansion exponent.
Several examples illustrate that the stability indices and expansion
exponents of quasi-isolated systems are not only asymptotically positive but
are asymptotically increasing. This means that the divergence of dynamical
trajectories and the expansion of phase volume at large time occurs with
acceleration. Such a strongly irreversible evolution of quasi-isolated
systems explains the irreversibility of time.

\end{abstract}

\vskip 1cm


\vskip 5mm


\newpage

The evolution of all real physical systems in nature is, as is well known,
irreversible. But the majority of physical laws is formulated as
time-reversible dynamical equations. The problem of obtaining irreversible
evolution of macroscopic systems from reversible deterministic dynamics
goes back to Boltzmann [1] who connected irreversibility with monotonic
growth of entropy in an isolated system evolving toward equilibrium. The
irreversibility of time in quantum systems was attributed by von Neumann
[2] to the measurement process. In the frame of a short communication, it
is impossible to give a detailed description of all approaches to and
nuances of the old and widely discussed problem of time irreversibility. A
good survey of the problem can be found in recent reviews [3,4] and a thorough
description of its history in book [5].

The approach, relating the irreversibility of the evolution laws of nature
with the increase of entropy in isolated systems tending to equilibrium,
has some weak points. First of all, there is no effective definition of
increasing entropy for strongly nonequilibrium systems [6]. Then, internal
chaotic dynamics in a real isolated system often does not provide finite
relaxation time to equilibrium or fast decay of fluctuations, but instead
displays a property of persistence of nonequilibrium [4]. Also, the
irreversibility of natural evolution laws is not the privilege of only
statistical systems characterized by entropy, even if the latter could be 
well defined. But it is rather a general law pertinent to all realistic 
systems, whether statistical or mechanical.

An important fact is that isolated systems as such are merely an abstraction.
No real system can be absolutely isolated from its surroundings [7--12]. Even
more, from the practical point of view, the concept of an isolated system is
logically self-contradictory, since to realize the isolation, one needs to
employ technical devices acting on the system; and to ensure that the latter
is kept isolated, one must apply measuring instruments perturbing the system
[13,14]. All real physical systems can be not more than quasi-isolated, that
is, almost isolated but, anyway, subject to the action of uncontrollable random
noise from their environment. Quasi-isolated systems are, in other words,
quasi-open and as such, even being in a seemingly steady state, can experience,
during their lifetime, large nonequilibrium fluctuations of entropy, density or
temperature [15--17].

An accurate mathematical treatment of quasi-isolated systems has recently
been advanced [18], from where it follows that these systems are, in
general, stochastically unstable. In the present communication, this result
is essentially strengthened in several aspects. In addition to the stability
index, describing the divergence of trajectories of quasi-isolated systems
[18], we introduce here the expansion exponent, for stochastic systems,
characterizing the expansion (or contraction) of the average phase-space
volume. The phase-volume expansion is, generally, a stronger characteristic
than the trajectory divergence. A system may be unstable, though its phase
volume be conserved, as it happens, e.g., for Hamiltonian systems [19]. But
if the phase volume expands, the system is certainly unstable. It turns 
out that in quasi-isolated systems at large time not only trajectories
diverge but also the phase volume expands. Moreover, the stability index and
the expansion exponent not merely become positive at large time but even
are increasing, which implies that the trajectory divergence and phase-volume
expansion proceed with acceleration. This essentially irreversible asymptotic
behaviour of quasi-isolated systems fixes the direction of time arrow. The
increase of time may be connected with the increase of the expansion exponent.
The strongly irreversible asymptotic evolution of quasi-isolated systems,
and hence of all real systems, can be interpreted as the cause of time
irreversibility.

In order to avoid misunderstanding, it is worth emphasizing the principal
points of the approach, presented in this paper, which distinguishes it
from all previous publications treating the irreversibility of time. A
detailed discussion of the standard ways of considering the arrow of time,
related to thermodynamic notions, can be found in Refs. [3--6,17--19].

\vskip 2mm

{\bf 1}. First of all, the consideration below is based on studying the
behaviour of {\it dynamical systems} but not that of statistical or
thermodynamic systems. For the former, contrary to the latter, there is no
need of dealing with entropy, entropy production, and other thermodynamic
characteristics, as well as with the second law of thermodynamics. This
different point of view looks more natural and general, since the
irreversibility of time is the property common not solely for thermodynamic
systems but for any real systems of this world.

\vskip 2mm

{\bf 2}. A pivotal concept, employed in the paper, is that of {\it
quasi-isolated systems}. Such a system is subject to the influence of {\it
infinitesimally weak uncontrollable} random perturbations, so that at each 
instant of time the zero-noise limit yields a time-reversible system. The 
irreversibility comes into play not merely because of {\it negligibly small} 
stochastic perturbations but due to the {\it noncommutativity} of the 
zero-noise and large-time limits.

\vskip 2mm

{\bf 3}. An important notion, the approach is based on, is the {\it local
expansion exponent} $X(t)$. For deterministic systems, there exists the
limit
$$
\lim_{t\rightarrow\infty}\; \frac{d}{dt}\; X(t) = \sum_n \lambda_n \; ,
$$
in which the right-hand side is the sum of all Lyapunov exponents. However,
for stochastic systems, such a relation in general does not exist, and
even more, the Lyapunov exponents as such often are not well defined
[23]. In addition, for our purpose of analysing the noncommutativity of
different limits, we need to consider the local temporal values of $X(t)$,
but not simply the large-time limit. This is why the usage of the local
expansion exponent is of principal importance.

The evolution of physical systems is usually described by partial differential
equations. Let $x\in\Bbb{D}\subset\Bbb{R}^d$ be a $d$-dimensional set of
continuous variables on a domain $\Bbb{D}$ and $t\in\Bbb{R}_+\equiv[0,\infty)$
denote time. Let a discrete index $i\in\Bbb{N}_+\equiv\{ 1,2,\ldots\}$
enumerate dynamical states. In what follows, we employ the matrix notation
[18,20] treating the pair $\{ i,x\}$ as a point in the label space $\Bbb{N}_+
\times\Bbb{D}$. Then the stochastic field $\xi(t)=[\xi_i(x,t)]$ is a column
with respect to the multi-index $\{i,x\}$. The dynamical state
$y(\xi,t)=[y_i(x,\xi,t)]$ is also a column with respect to $\{ i,x\}$, as is
the vector field $v(y,\xi,t)=[v_i(x,y,\xi,t)]$. The evolution of a dynamical
system, subject to the action of noise, is presented by a stochastic
differential equation
\be
\label{1}
\frac{d}{dt}\; y(\xi,t) = v(y,\xi,t) \; ,
\ee
with an initial condition $y(\xi,0)=y(0)$ and given boundary conditions.
Here and everywhere below, the stochastic differential equations are
understood in the sense of Stratonovich [21]. As the main measurable
quantity, we consider the average trajectory
\be
\label{2}
y(t)= [y_i(x,t)] \equiv \; \ll y(\xi,t)\gg \; ,
\ee
with the averaging accomplished over the stochastic field.

Important characteristics of motion are the stochastic multiplier matrix
$\hat M(\xi,t)=[M_{ij}(x,x',\xi,t)]$, with the elements
\be
\label{3}
M_{ij}(x,x',\xi,t) \equiv \frac{\dlt y_i(x,\xi,t)}{\dlt y_j(x',0)} \; ,
\ee
and the stochastic Jacobian matrix $\hat J(\xi,t)=[J_{ij}(x,x',\xi,t)]$,
whose elements are
\be
\label{4}
J_{ij}(x,x',\xi,t) \equiv
\frac{\dlt v_i(x,y,\xi,t)}{\dlt y_j(x',\xi,t)}\; .
\ee
These matrices are connected through the equation
\be
\label{5}
\frac{d}{dt}\; \hat M(\xi,t) = \hat J(\xi,t)\; \hat M(\xi,t) \; ,
\ee
following from the variation of Eq. (1). Equation (5) is to be
complimented with the initial condition $\hat M(\xi,0)=\hat 1=[\dlt_{ij}\;
\dlt(x-x')]$ and the boundary conditions resulting from the variation of
those for the dynamical state (see details in Ref. [18]).

The stability of the system is characterized by considering the behaviour
of the trajectory deviation $||\dlt y(t)||$ caused by an infinitesimal
variation of initial conditions. Here the Hermitian vector norm
$$
||y(t)|| \equiv \left [ \sum_i \int y_i^*(x,t)\; y_i(x,t)\; dx \right ]^{1/2}
$$
is assumed. The {\it local stability index} is defined [18] as
\be
\label{6}
\sgm(t) \equiv \ln\; \sup_{\dlt y(0)}\;
\frac{||\dlt y(t)||}{||\dlt y(0)||} \; .
\ee
This shows the trajectory deviation $||\dlt y(t)||\sim||\dlt y(0)||^{\sgm(t)}$
at a given time $t$. The stability index (6) can be expressed [18] through
the multiplier matrix as
\be
\label{7}
\sgm(t) = \ln|| \ll \hat M(\xi,t)\gg || \; ,
\ee
where the Hermitian norm of the matrix is meant.

To describe the behaviour of the phase-space volume, we need, first, to
define [22] the {\it elementary phase volume}
\be
\label{8}
\dlt\Gm(t) \equiv \prod_i\; \prod_x \dlt y_i(x,t) \; ,
\ee
where the continuous product over $x$ is specified in Ref. [15]. Now, we
introduce the {\it local expansion exponent}
\be
\label{9}
X(t) \equiv \ln \left | \left | \frac{\dlt\Gm(t)}{\dlt\Gm(0)} \right |
\right | \; ,
\ee
which determines the temporal behaviour of the phase volume $||\dlt\Gm(t)||
\sim||\dlt\Gm(0)||e^{X(t)}$. When $X(t)<0$, the phase volume at time $t$ 
contracts; if $X(t)=0$, the volume is preserved; and when $X(t)>0$, it 
expands. The expansion exponent (9) can be related to the multiplier 
matrix as $X(t)={\rm Tr}\hat L(t)$, with
$$
\hat L(t) =\left [\ln |\ll M_{ij}(x,x',\xi,t)\gg | \right ] \; .
$$
For a matrix $\hat A$, we have the identity ${\rm Tr}\ln A=\ln\det\hat A$.
Employing the latter, we obtain
\be
\label{10}
X(t) = \ln|\det\ll \hat M(\xi,t)\gg | \; .
\ee
Equations (7) and (10) can be simplified invoking the diagonal representation
for the multiplier matrix, when
\be
\label{11}
M_{mn}(\xi,t) =\dlt_{mn}\; \mu_n(\xi,t)\; ,
\ee
with $m$ and $n$ being the appropriate multi-indices [18].

Quasi-isolated systems are those that are subject to the action of a weak
stochastic noise modelling the random uncontrollable perturbations from the 
environment [18]. To have the possibility for varying the noise amplitude, it 
is convenient to affix to the stochastic field $\xi(t)$ an explicit factor 
$\al$ whose value could be regulated, for which we change $\xi(t)$ by 
$\al\xi(t)$. Then the stability index (7) and expansion exponent (10) acquire 
the dependence on $\al$. For instance, in the diagonal representation (11), 
we get the stability index
\be
\label{12}
\sgm_\al(t) =\sup_n\; \ln |\ll \mu_n(\al\xi,t)\gg |
\ee
and the expansion exponent
\be
\label{13}
X_\al(t) = \sum_n \ln |\ll \mu_n(\al\xi,t)\gg | \; .
\ee

To illustrate explicitly the similarities and differences between the
stability index and expansion exponent, let us consider several examples
of stochastic systems, for which the multiplier matrix can be calculated
exactly. To simplify calculations, we treat the stochastic field $\xi(t)$
as a scalar white noise centered at zero, with the averages
\be
\label{14}
\ll\xi(t)\gg\; = 0 \; , \qquad \ll\xi(t)\;\xi(t')\gg\; =
2\gm\; \dlt(t-t') \; ,
\ee
where $\gm>0$. The technique of such calculations was explained in detail
in Ref. [18]. Therefore, here we will not write down intermediate
manipulations but shall present only the results.

An oscillatory process in stochastic background is described by the equation
\be
\label{15}
\frac{dy}{dt} = (i\om + \al\xi) \; y \; ,
\ee
where $y=y(\al\xi,t)$, $\xi=\xi(t)$, and $\om$ is a real frequency. If
$\al$ is real-valued, then the stochastic term corresponds to the
attenuation-generation noise. For the local multiplier, we get
$$
\mu(\al\xi,t) =\exp\left\{ i\om t +
\al \int_0^t \xi(t')\; dt' \right \} \; ,
$$
which yields
\be
\label{16}
\sgm_\al(t) =  X_\al(t) = \al^2\gm t \; .
\ee

The divergence of trajectories for stochastic systems is not necessarily
exponential but can be of algebraic law [23]. For this to occur, one has
to consider slightly more complicated equations or a coloured noise. Here
we give an explicit example of the arising algebraic equation with white
noise. Consider the process
\be
\label{17}
\frac{dy}{dt} = v_0 + \frac{\al\xi}{\sqrt{1+t}} \; y \; ,
\ee
where $v_0=v_0(t)$ does not include $y$ and $\al$ is again real. The
local multiplier is
$$
\mu(\al\xi,t) =\exp\left\{ \al \int_0^t
\frac{\xi(t')}{\sqrt{1+t'}}\; dt' \right \} \; .
$$
The stochastic averaging results in the power law
\be
\label{18}
\ll \mu(\al\xi,t)\gg = (1+t)^{\al^2\gm} \; ,
\ee
characterizing the algebraic divergence of the system trajectories. The
stability index and expansion exponent become
\be
\label{19}
\sgm_\al(t) = X_\al(t) = \al^2\gm \ln (1+t) \; .
\ee

In the above examples, the stability index and expansion exponent coincide
because of the one-dimensionality of the considered dynamical systems.
The situation is different for higher-dimensional systems. As an example,
let us examine the stochastic Schr\"odinger equation
\be
\label{20}
\frac{\partial\psi}{\partial t} =
\left ( -i\; H +\al\xi\right )\; \psi \; ,
\ee
in which $\psi=\psi(\br,\al\xi,t)$; $\br\in\Bbb{R}^3$; $H=H(\br)$ is
a Hamiltonian, and the Planck constant $\hbar\equiv 1$. The stochastic
term, with a real-valued $\al$, imitates an attenuation-generation
influenced by a random surrounding. The multiplier matrix, in the
diagonal representation (11), possesses the elements
$$
\mu_n(\al\xi,t) =\exp\left\{ -i\; E_n\; t +\al
\int_0^t \xi(t')\; dt' \right \} \; ,
$$
where $E_n$ are the eigenvalues of the Hamiltonian, defined by the
stationary problem $H\psi_n= E_n\psi_n$, with a multi-index $n$
enumerating quantum states. Denoting the total number of states as
$N=\sum_n 1$, we find the stability index and expansion exponent,
\be
\label{21}
\sgm_\al(t) =\al^2\gm t \; , \qquad X_\al(t) =\al^2 N\gm t \; ,
\ee
which differ by the factor $N$.

As another case of an infinite-dimensional dynamical system, let us treat
the stochastic diffusion equation
\be
\label{22}
\frac{\partial y}{\partial t} = (D +\al\xi)\;
\frac{\partial^2 y}{\partial x^2} \; ,
\ee
in which $y=y(x,\al\xi,t)$; $x\in[0,1]$; $D>0$. Since the diffusion
constant $D$ plays the role of an attenuation parameter, the stochastic
term with a real $\al$ again models the relaxation-generation impact of
a random environment. Besides an initial condition, Eq. (22) is to be
supplemented by the boundary conditions, which we take in the form
$y(0,\al\xi,t)=b_0$ and $y(1,\al\xi,t)=b_1$, where $b_0$ and $b_1$ are
constants. The diagonal representation for the multiplier matrix gives
$$
\mu_n(\al\xi,t) =\exp\left\{ -D(\pi n)^2 t - \al(\pi n)^2
\int_0^t \xi(t')\; dt'\right \} \; ,
$$
where $n=1,2,\ldots, N$, with $N\rightarrow\infty$. From here, we find
the stability index (12) and expansion exponent (13),
$$
\sgm_\al(t)\simeq - \dlt_{\al 0}\; D\pi^2 t + (1 -\dlt_{\al 0} )
\al^2 (\pi N)^4 \gm t \; ,
$$
\be
\label{23}
X_\al(t) \simeq -\; \frac{\pi^2}{3}\; N^3 Dt + \al^2\;
\frac{\pi^4}{5}\; N^5 \gm t,
\ee
whose forms are rather different.

Analysing the cases, considered above, resulting in Eqs. (16), (19), (21),
and (23), we see the following. The limits $\al\ra 0$ and $t\ra\infty$
do not commute for the stability index $\sgm_\al(t)$, hence these
quasi-isolated systems are stochastically unstable, as discussed in Ref.
[18]. In addition to this, we find that these limits do not commute also
for the expansion exponent $X_\al(t)$. Moreover, for the commutator of
the limits, we have
\be
\label{24}
[\lim_{\al\ra 0}, \; \lim_{t\ra\infty}]\; \sgm_\al(t) = \infty\; , \qquad
[\lim_{\al\ra 0},\; \lim_{t\ra\infty}]\; X_\al(t) = \infty \; .
\ee
This happens because the deterministic systems, obtained when switching
off stochastic terms by setting $\al\ra 0$, are either stable and
phase-volume contracting or neutrally stable and phase-volume preserving,
while in the presence of arbitrary small stochastic terms, the quasi-isolated
systems become unstable and phase-volume expanding, such that
\be
\label{25}
\sgm_\al(t) \ra \infty\; , \qquad X_\al(t) \ra \infty \qquad
(\al\neq 0, \; t\ra\infty)\; .
\ee
Repeating the same arguments as in Ref. [18], it can be inferred that the
asymptotic property (25) holds true for quasi-isolated systems in general.

It is important to emphasize that such an instability, as is described above,
certainly happens not for any weak noise. One could easily find examples
when a weak external noise, vice versa, would suppress chaotic behaviour, 
thus, stabilizing the considered system. This may occur, e.g., for some 
cases of additive noise. As is explained in Ref. [18], the existence of
{\it multiplicative} noise is a paramount requirement for the appearance
of stochastic instability. It is also clear that not any multiplicative 
noise would produce instability. But we must always keep in mind that the 
central feature of a quasi-isolated system is the presence of noise that, 
though being weak, is also {\it uncontrollable}. The latter implies that,
analysing the stochastic stability of a given system, we are obliged to 
check the stability with respect to all admissible types of noise and to 
select among those the worst situation, that is, the noise whish results 
in maximal instability. This maximization of instability is what is meant 
when terming the noise as uncontrollable. And the existence of 
infinitesimally weak {\it uncontrollable perturbations} is in the heart 
of the definition of {\it quasi-isolated systems} [18]. This principal 
feature has always been silently assumed when chosing the type of 
perturbations for the examples of the present paper, so that the most 
unstable case has always been treated.  By definition, uncontrollable 
includes the worst possible, which in the present context means the least
stable.

In this way, quasi-isolated systems not only are stochastically unstable
but their phase-space volume asymptotically expands. What is more, both
the stability index and the expansion exponent are increasing at large
time. This means that the divergence of trajectories and phase-volume
expansion occur with acceleration. To imagine this picture, one may keep
in mind the accelerated expansion of Universe [24,25]. The asymptotic
increase of the stability index and expansion exponent for quasi-isolated
systems shows that the evolution of the latter is essentially irreversible.
Actually, the increase of the stability index, expansion exponent, and time
occurs, according to the law (25), simultaneously. Since all real-world
systems are not more than quasi-isolated, the irreversibility of time is
a general law of nature.

\vskip 5mm

{\bf Acknowledgement}

\vskip 2mm

I am very grateful to Y.E. Lozovik for constructive discussions and highly 
useful advice.

I appreciate financial support from the S\~ao Paulo State Research
Foundation, Brazil and Bogolubov-Infeld International Program, Poland.

\end{document}